\input{aipcheck}

\documentclass[final] {aipproc}

\layoutstyle{6x9}

\begin{document}

\title{New Statistical Results on the Angular Distribution of Gamma-Ray Bursts}

\classification{ 01.30.Cs,  95.55. Ka,  95.85.Pw, 98.70.Rz}
\keywords      {gamma ray burst, factor analysis}

\author{ Lajos G. Bal\'azs}{ address={ Konkoly Observatory, H-1525 Budapest, POB 67, Hungary}}

\author{Istv\'an Horv\'ath}{ address={ Dept. of Physics, Bolyai Military
University, H-1581 Budapest, POB 15, Hungary}}

\author{Roland Vavrek}{ address={ ESA/ESAC P.O. Box 50727
Villafranca del Castillo, 28080 Madrid, Spain}}

\author{Zsolt Bagoly}{ address={Dept. of Phys. of Complex Systems, E\" otv\"
os Univ., H-1117 Budapest, P\'azm\'any P. s. 1/A, Hungary}} 

\author{Attila M\'esz\'aros}{ address={ Astron. Inst. of the Charles
University, V Hole\v{s}ovi\v{c}k\'ach 2, CZ-180 00 Prague 8, Czech Republic}}

\begin{abstract}
We presented the results of several statistical
tests of the randomness in the angular sky-distribution of gamma-ray bursts
in BATSE Catalog. Thirteen different tests were presented based on Voronoi
tesselation, Minimal spanning tree and Multifractal spectrum for five classes
(short1, short2, intermediate, long1, long2) of gamma-ray bursts, separately.
The long1 and long2 classes are distributed randomly. The intermediate subclass, in accordance with
the earlier results of the authors, is distributed non-randomly. Concerning the
short subclass earlier statistical tests also suggested some departure from the
random distribution, but not on a high enough confidence level. The new
tests presented in this article suggest also non-randomness here.

\end{abstract}

\maketitle


\section{Introduction}
	There are increasing evidences that all the GRBs do not represent a physically homogeneous group 
(Kouveliotou et al. 1993, horv\'ath 1998, bal\'azs et al. 2003, Hakkila et al. 2003, horv\'ath et al. 2006). Hence, it is worth investigating that the physically different subgroups are also different in their angular distributions. In the last years the authors provided 
(bal\'azs et al. 1998, bal\'azs et al. 1999, M\'esz\'aros et al. 2000) several different tests probing the intrinsic isotropy in the angular sky-distribution of GRBs collected in BATSE Catalogs. 
One may conclude the results of these studies: 
	A. The long subgroup  seems to be distributed isotropically (see also Briggs 1993); 
	B. The intermediate subgroup (horv\'ath 2002, horv\'ath et al. 2006)
is distributed anisotropically on the 96-97\% significance level; 
	C. For the short subgroup the assumption of isotropy is rejected only on the 92\% significance level; 
	D. The long and the short subclasses, respectively, are 
distributed differently on the 99.3\% significance level.  
	Independently (Litvin et al. 2001),  confirmed the results A., 
B. and C. with one essential difference: for the intermediate 
subclass a much higher - namely 99.89\% - significance level of 
anisotropy is claimed. Again, the short subgroup is found to 
be "suspicious", but only on the 85-95\% significance level. 
	In this paper, similarly to the previous studies, 
the intrinsic randomness is tested; this means that the 
non-uniform sky-exposure function of BATSE instrument was considered.

\section{Mathematical summary and the test-variables}
\label{mat}

The randomness of the point field on the sphere
 can be tested with respect to different criteria. In the following we 
defined several test-variables.

{\it Voronoi tesselation (VT).}
The Voronoi diagram - also known as Dirichlet tesselation or Thiessen 
polygons - is a fundamental structure in computational geometry 
(Voronoi 1908, Stoyan \& Stoyan 1994). Generally, this diagram 
provides a partition of a point pattern 
according to its spatial structure. 
	Assume that there are N points (N >> 1) scattered on a sphere 
surface with an unit radius. 
The Voronoi cell of a point is the region of the sphere surface 
consisting of points which are closer to this given point than 
to any other ones of the sphere. This cell forms a polygon on 
this sphere. Every such cell has its area (A) given in steradians, 
perimeter (P) given by the length of boundary 
(one great circle of the boundary curve is called also as "chord"), 
number of vertices ($N_v$) given by an integer positive number, 
and by the inner angles. This method is completely non-parametric, and therefore may be sensitive for various point pattern structures in the different subclasses of GRBs.

Any of the four quantities characterizing the Voronoi cell
can be used as test-variables or even some of their
combinations, too. We defined the following test-variables: 
1, Cell area  $ A $; 
2, Cell vertex (edge)   $ N_v $; 
3, Cell chords    $ C $; 
4, Inner angle  $ \alpha_i$; 
5, Round factor (RF) average $RF_{av}=\overline{4\pi A / P
}$; 
6, Round factor (RF) homogeneity  $1-\frac{\sigma
(RF_{av})}{RF_{av}}$; 
7, Shape factor   $ A/P^2$; 
8, Modal factor   $\sigma (\alpha_i) / N_v$; 
9, The so-called "AD factor"
defined as $AD = 1 - (1 - \sigma (A) / \langle A \rangle )^{-1}$.

\begin{table}
\centering
  \caption{Tested samples of BATSE GRBs.}
  \label{test}
  \begin{tabular}{@{}lllc}
  \hline
  Sample & Duration  & Peak flux  & Number  \\
         &    [$s$]    &  [$photons$ $cm^{-2}$$s^{-1}$] & of GRBs   \\
  \hline
  Short1  &  $T_{90}<2$ s & $0.65 < P_{256} < 2$ & 261 \\
  Short2  &  $T_{90}<2$ s & $0.65 < P_{256}$     & 406 \\
  Intermediate   & $2\;s < T_{90} < 10\;s $ & $0.65 < P_{256}$ & 253 \\
  Long1   &  $T_{90}>2$ s & $0.65 < P_{256} < 2$ & 676 \\
  Long2   &  $T_{90}>10$ s & $0.65 < P_{256}$    & 966 \\
  \hline
  \end{tabular}
\end{table}

{\it Minimal spanning tree (MST).}
 Contrary to VT, this method considers the distances (edges) 
among the points (vertices). A spanning tree is a system of 
lines connecting all the points without any loops. 
The minimal spanning tree (MST) is a system of connecting lines, 
where the sum of the lengths is minimal among all the possible 
connections between the points  (Prim 1957).  
The statistics of the lengths and the MST angles between the 
edges at the vertices  can be used for testing the randomness 
of the point pattern. 
To characterize the stochastic properties of a point patters we
use three quantities obtained from a MST: 
1, Variance of the MST edge-length
$\sigma(L_{MST})$; 
2, Mean MST edge-length   $L_{MST}$; 
3, Mean  angle between edges   \ $\alpha_{MST}$.

{\it Multifractal spectrum} is the third method which was used.
Here the only used variable is the  $f(\alpha)$ multifractal spectrum, which 
 is a sensitive tool for testing the non-randomness of a point pattern.

\begin{table}
\centering \caption{Calculated significance levels for the 13
test-variables and the five samples. A significance  greater than
95\% is put in bold face.}\label{testtab}
\begin{tabular}{llcccccc}
\hline
 Name &  var & short1  & short2 & interm. &  long1  &  long2\\
 \hline
Cell area  &  $ A $ &  36.82  &   29.85  &   94.53  &  79.60 &   82.59\\
Cell vertex (edge)  & $ N_v $  &  36.82  &   87.06  &  \ 2.99  &  26.87 &  \ 7.96\\
Cell chords  &  $ C $  &  47.26  &   52.24  &   18.91  &  84.58 &   54.23\\
Inner angle  &  $ \alpha_i $  &  {\bf 96.52}  &   21.39  &   87.56  &  37.81 &   63.18\\
RF average &  $\overline{4\pi A / P }$  &  65.17  &  {\bf 99.98}  &   33.83  &  10.95 &   86.07 \\
RF homogeneity & $1-\frac{\sigma (RF_{av})}{RF_{av}}$   &  19.90  &   24.38  &   58.71  &  55.72 &   32.84 \\
Shape factor  &  $ A/P^2 $  &  91.04  &   94.03  &   90.05  &  55.22 &   63.68\\
Modal factor &  $\sigma (\alpha_i) / N_v$  &  {\bf 97.51}  &  \ 1.99  &  \ 7.46  &  56.22 &  \ 8.96\\
AD factor &  $1- \bigl(1- \frac{\sigma (A)} { \langle A \rangle}\bigr)^{-1}$   &  32.84  &   25.37  &   11.44  &  {\bf 95.52} &   52.74 \\
MST variance  & $\sigma(L_{MST})$     & 52.74  &   38.31  &   22.39  &  13.93 &   59.70\\
MST average  &  $L_{MST} $     & {\bf 97.51}  &  \ 7.46  &   89.05  &  56.72 &  \ 8.96\\
MST angle   & \ $\alpha_{MST}$     & 85.07  &   14.43  &   36.82  &  73.63 &   60.70\\
MFR spectra &  $f(\alpha)$      & {\bf 95.52}  &   {\bf 96.02}  &   {\bf 98.01}  &  73.63 &   36.32\\
\hline Binomial test &  & {\bf99.79} &
{\bf99.74} &
77.00 & 55.13 & -  \\
 \hline
Squared Euclidean & distance &  {\bf 99.90} & {\bf 99.98}  & {\bf 98.51} & 93.03  &   36.81       \\
\hline
\end{tabular}
\end{table}

\section{Results}

Completing 200 simulations in all of the subsamples (for them see  Table 1.) we get 
 a 13D sample representing the joint probability distribution of the 13 test variables. Using a suitable chosen measure 
of distance of the points from the sample mean we can get a stochastic variable characterizing 
the deviation of the simulated points from the mean only by chance. An obvious choice would be the squared Euclidean distance.

	In case of a Gaussian distribution with unit variances and 
without correlations this would resulted in a  $\chi^2 $ distribution of 13 degree of freedom. But the test variables in our case 
are correlated and have different scales. 
	Factor analysis (FA) is a suitable way to represent the correlated observed variables with fewer non-correlated variables 
of less in number (Wallet \& Dussert 1998).  
The number of non-correlated variables, 
k, can be constrained by k < 8.377 in our case for n = 13. Hence, we retained 8 non-correlated variables.

\begin{figure}
\centering
  \includegraphics[width=91mm]{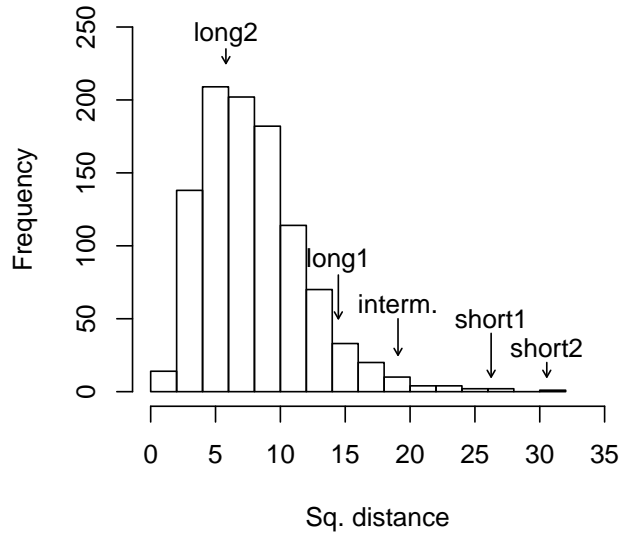}
  \caption{Distribution of the Euclidean distances of the simulated samples
  from the stochastic mean of the variables in the 13D parameter
  space. There are altogether 1000 simulated points. Full line marks
  a $\chi^2$ distribution of 8 degree of freedom, normalized to the sample size.
  The distances of the BATSE samples
  are also indicated. The departures of samples $"short1"$ and
  $"short2"$ exceed all those of the simulated points. The
  probabilities, that these deviations are non-random, equal
  99.9\% and 99.98\%.
}
  \label{hist}
\end{figure}

Out of the 13 test-variables only the multifractal spectrum gave significant (>95\%) deviation from the simulated sample  
in more than one group. The BATSE samples, however, were different in the number of test-variables giving positive signal 
(>95\%) and in the level of significance. Among the tested samples short1 experiences four 
(96.5\%, 97.5\%, 97.5\%, 95.5\%), short2 two (99.98\%, 96.02\%), intermediate one (98.0\%), long1 one (95.5\%) 
and long2 no variables with >95\% significance (see Table 2.). 
Calculating the joint significance level we assumed that they can be represented as a linear combination of non-correlated hidden factors of less in number. 
We obtained k=8 as the number of hidden factors. Then we computed the distribution of 
the squared Euclidean distances from the mean of the simulated 
variables. Comparing the distribution of the squared distances 
of the simulated  with  the BATSE samples we concluded that the 
short1, short2 and intermediate groups deviate significantly 
(99.90\%, 99.98\% and 98.51\%) from the fully randomness but 
it is not the case at the long samples (see Fig. 1.).

{\it Acknowledgments:}
This study was supported by OTKA grant No. T048870 and 75072 and
Res. Program MSM0021620860  and GAUK grant No.46307 (A.M.).


\begin{thebibliography}{99}


\bibitem[\protect\citeauthoryear{Bal\'azs et al.}{1998}]{ba98} Bal\'azs L.G.,
M\'esz\'aros A., Horv\'ath I., 1998, A\&A, 339, 1

\bibitem[\protect\citeauthoryear{Bal\'azs et al.}{1999}]{ba99}
Bal\'azs L.G., M\'esz\'aros A., Horv\'ath I., Vavrek R., 1999,
A\&A Suppl., 138, 417

\bibitem[\protect\citeauthoryear{Bal\'azs et al.}{2003}]{bal03}
Bal\'azs L.G., Bagoly Z., Horv\'ath I., M\'esz\'aros A.,
M\'esz\'aros P., 2003, A\&A, 401, 129


\bibitem[\protect\citeauthoryear{Briggs}{1993}]{brig93}
Briggs M., 1993, ApJ, 407, 126


\bibitem[\protect\citeauthoryear{Hakkila et al.} {2003}]{hak03} Hakkila, J., et al.,
2003, ApJ, 582, 320

\bibitem[\protect\citeauthoryear{Horv\'ath}{1998}]{ho98}
Horv\'ath I., 1998, ApJ, 508, 757

\bibitem[\protect\citeauthoryear{Horv\'ath}{2002}]{ho02}
Horv\'ath I., 2002, A\&A, 392, 791

\bibitem[\protect\citeauthoryear{Horv\'ath et al.}{2006}]{ho06}
Horv\'ath I.,   Bal\'azs L.G.,  Bagoly Z., Ryde F., M\'esz\'aros
A., 2006, A\&A, 447, 23

\bibitem[\protect\citeauthoryear{Kouveliotou et al.}{1993}]{kou93}
Kouveliotou, C., et al., 1993, ApJ, 413, L101

\bibitem[\protect\citeauthoryear{Litvin et al.}{2001}]{li01}
Litvin V.F., Matveev S.A., Mamedov S.V., Orlov V.V., 2001, Pis'ma
v Astron. Zhurnal, 27, 489

\bibitem[\protect\citeauthoryear{M\'esz\'aros et al.}{2000}]{me00b}
M\'esz\'aros A., Bagoly Z., Horv\'ath I., Bal\'azs L.G., Vavrek
R., 2000b, ApJ, 539, 98

\bibitem[\protect\citeauthoryear{Prim}{1957}]{prim57}
Prim R.C., 1957, Bell Syst. Techn. Journ., 36, 1389

\bibitem[\protect\citeauthoryear{Stoyan \& Stoyan}{1994}]{stoy}
Stoyan D., Stoyan H., 1994, Fractals, Random Shapes and Point
Fields, Wiley J. \& Sons, New York

\bibitem[\protect\citeauthoryear{Voronoi}{1908}]{vor}
Voronoi G., 1908, J. Reine Angew. Math., 134, 198

\bibitem[\protect\citeauthoryear{Wallet \& Dussert}{1998}]{wadu98}
Wallet, F., Dussert C., 1998, Europhys. Let., 42, 493

\end{thebibliography}
\end{document}